\documentclass[twocolumn,apl,amsmath,amssymb,showpacs]{revtex4-2}
\usepackage{epsf}      
\usepackage{graphicx}
\usepackage{color}
\usepackage{soul}
\usepackage{gensymb}
\usepackage{sidecap}
\usepackage{amsmath}
\usepackage{mathtools}
 \usepackage{multirow}
\usepackage[hidelinks,colorlinks=true,linkcolor=blue,citecolor=blue]{hyperref}
\DeclareUnicodeCharacter{2212}{\textendash}

\hypersetup{
  colorlinks=false,
  colorlinks=true,
  citecolor=blue,
  linkcolor=blue,
  urlcolor=blue}

\begin{document}
\title{Magnetic, transport and electronic properties of Ni$_2$FeAl Heusler alloy nanoparticles: Experimental and theoretical investigation}

\author{Priyanka Yadav}
\affiliation{Department of Physics, Indian Institute of Technology Delhi, Hauz Khas, New Delhi-110016, India}
\author{Mohd Zeeshan}
\affiliation{Department of Physics, Indian Institute of Technology Delhi, Hauz Khas, New Delhi-110016, India}
\author{Brajesh K. Mani}
\email{bkmani@physics.iitd.ac.in}
\affiliation{Department of Physics, Indian Institute of Technology Delhi, Hauz Khas, New Delhi-110016, India}
\author{Rajendra S. Dhaka}
\email{rsdhaka@physics.iitd.ac.in}
\affiliation{Department of Physics, Indian Institute of Technology Delhi, Hauz Khas, New Delhi-110016, India}

\date{\today}      

\begin{abstract}
We present a comprehensive investigation of structural, magnetic and transport properties of Ni$_2$FeAl Heusler alloy nanoparticles (NPs) synthesized via template-less chemical route. The NPs 
exhibit high saturation magnetization of 3.02 $\mu_ {\rm B}$/f.u. at 5~K,
large magnetic anisotropy of 0.238 MJ/m$^3$, and a Curie temperature of 874~K. Magnetocaloric analysis reveals a magnetic entropy change of 3.1 J.kg$^{-1}$K$^{-1}$ at 70 kOe. Low-temperature transport measurements show a weak resistivity upturn, following a $-T^{1/2}$ 
dependence, indicative of disorder-enhanced electron-electron interactions. 
First-principles calculations based on density functional theory yield a magneto-crystalline anisotropy energy of 0.987 MJ/m$^3$, consistent with experiment and demonstrate pronounced surface and finite-size effects through comparison of bulk and nanocluster geometries. The combination of 
high Curie temperature, sizable perpendicular magnetic anisotropy, 
and moderate spin polarization and magnetic entropy change make the Ni$_2$FeAl as promising candidate for various applications.
	
\end{abstract}
\maketitle

\section{Introduction}

Heusler alloys constitute a versatile class of materials having potential applications in thermoelectrics, magnetic refrigeration, spintronics and shape-memory 
technologies \cite{Krenke_PRB_2006, Krenke_Nat_2005, Wang_JAP_2007, Groot_PRL_1983, PY_JMMM_2025}. While their bulk behavior has been extensively characterized, increasing attention is currently being directed towards low-dimensional counterparts, where reduced symmetry and enhanced surface-to-volume ratios give rise to emergent electronic and magnetic phenomena. {\it Ab initio} studies have indicated that interfacial and finite-size effects critically influence phase stability, half-metallicity, and magnetic interactions in these systems, thereby motivating systematic investigations at the nanoscale \cite{Galanakis_JPCM_2002, Endo_JPDP_2011, Hashemifar_PRL_2005}. 
Concurrently, Heusler-based nanostructures including thin films, nanowires and 
nanoparticles, have been explored intensively for high-performance 
spintronic and magnetoresistive devices, \cite{Wang_PRB_2005, Bosu_PRB_2011, Galdun_ACSN_2018}, as well as for applications involving magnetic nanoparticles (NPs) in biomedical imaging, drug delivery, catalysis, nanoscale magnetic storage, and magnetic refrigeration \cite{Poorvi_EJIC_2024, Karim_Nanot_2022, GD_ACS_2021, PRL_2007, Sun_sci_2000, Billas_JMMM_97, Fabris_Nano_2019}. Within this landscape, Co-based Heusler nanoparticles are of particular interest owing to their high Curie temperatures, large magnetic moments, and potential half-metallicity \cite{Nehla_JALCOM_2019, Ahmad_JMMM_2019, Xu_AIP_2018}. Pronounced size- and disorder-driven magnetic effects have been reported: coercivity scaling in Co$_2$FeAl \cite{Ahmad_JMMM_2019}, enhanced magnetization in template-free Co$_2$FeGa  \cite{Nehla_JALCOM_2019} and secondary $\gamma$-phase-assisted tunability in Co$_2$NiGa  \cite{Wang_JMCC_2016}, reinforcing that synthesis-induced disorder and finite-size effects play a decisive role in governing magnetic functionality in nanoscale Heusler systems. 

Beyond Co-based systems, Mn- and Ni-based Heusler alloys have emerged as technologically versatile platforms. The Mn-based tetragonal Heusler alloys offer rare-earth-free permanent magnet behavior driven by strong spin--orbit coupling and enhanced magnetocrystalline anisotropy (MCA) \cite{Wollmann_PRB_2015}, whereas Ni-based variants serve as benchmark ferromagnetic shape-memory and magnetocaloric materials with martensitic transformations, tunable anisotropy, and substantial spin polarization—making them promising for multifunctional spintronic and actuation applications \cite{Wang_JAP_2007, Webster_PM_1984, Uijttewaal_PRL_2009}. This tunability is particularly relevant for spin-transfer-torque (STT) devices, where materials must simultaneously deliver high Curie temperature, sizable spin polarization, controlled anisotropy, and moderate magnetization \cite{Winterlik_AM_2013}. Although achieving low switching current and fast switching while retaining thermal stability is challenging, tetragonally distorted Heusler alloys, particularly Mn$_{3-x}$Ga ($x=0-1$) \cite{Wu_APL_2009}, address this requirement through intrinsic perpendicular magnetic anisotropy, enabling energy-efficient switching without sacrificing stability. Their rich structural and compositional tunability further allows decoupled control of saturation magnetization, magnetocrystalline anisotropy, and damping, providing a pathway to optimize performance across memory, logic, and spin-torque oscillator technologies \cite{Winterlik_AM_2013}.

Within this context, Ni$_2$FeAl remains as an intriguing yet under-explored candidate. Prior studies have reported intriguing low-temperature magnetic features: melt-spun ribbons reveal blocked super-paramagnetic clusters arising from intergranular interactions \cite{Zhang_JPCM_2007}, an unusual behavior for chemically ordered Heusler alloys \cite{Slebarski_PRB_2001, Slebarski_PRB_2002}. These observations have been interpreted in terms of nanoscale disorder-driven magnetic inhomogeneities and point to a critical gap in understanding the interplay of reduced dimensionality, magnetic disorder and spin-orbit coupling in Ni$_2$FeAl. Establishing quantitative structure–property relationships in Ni-based Heusler nanoparticles, especially in chemically disordered and low-dimensional regimes, is essential for evaluating their potential as scalable practical use. Therefore, considering these open questions, we undertake a detailed investigation of Ni$_2$FeAl nanoparticles, focusing on their structure, magnetic, electronic and transport properties.

In this work, we report a comprehensive study of single-domain Ni$_2$FeAl 
Heusler alloy nanoparticles synthesized via a template-free chemical approach. 
Structural and morphological analyses confirm their crystallization in the 
I4/$mmm$ phase. Magnetic measurements reveal a high 
saturation magnetization and significant uniaxial anisotropy, suggesting 
its potential for STT and PMA-driven applications. Low-temperature magnetotransport measurements exhibit a resistivity upturn with a characteristic $-T^{1/2}$ dependence, indicative of disorder-enhanced electron-electron interactions. First-principles calculations, incorporating spin-orbit coupling, yield a large magneto-crystalline anisotropy energy, consistent with experimental estimates. Additional simulations on bulk-like nanoclusters further underscore the role of surface and finite-size effects in modulating the electronic structure and 
magnetic properties. These findings establish Ni$_2$FeAl nanoparticles 
as promising candidate for next-generation nanospintronic, magnetoresistive, 
and energy-efficient magnetic technologies.

\begin{figure*}
\includegraphics[width=7in]{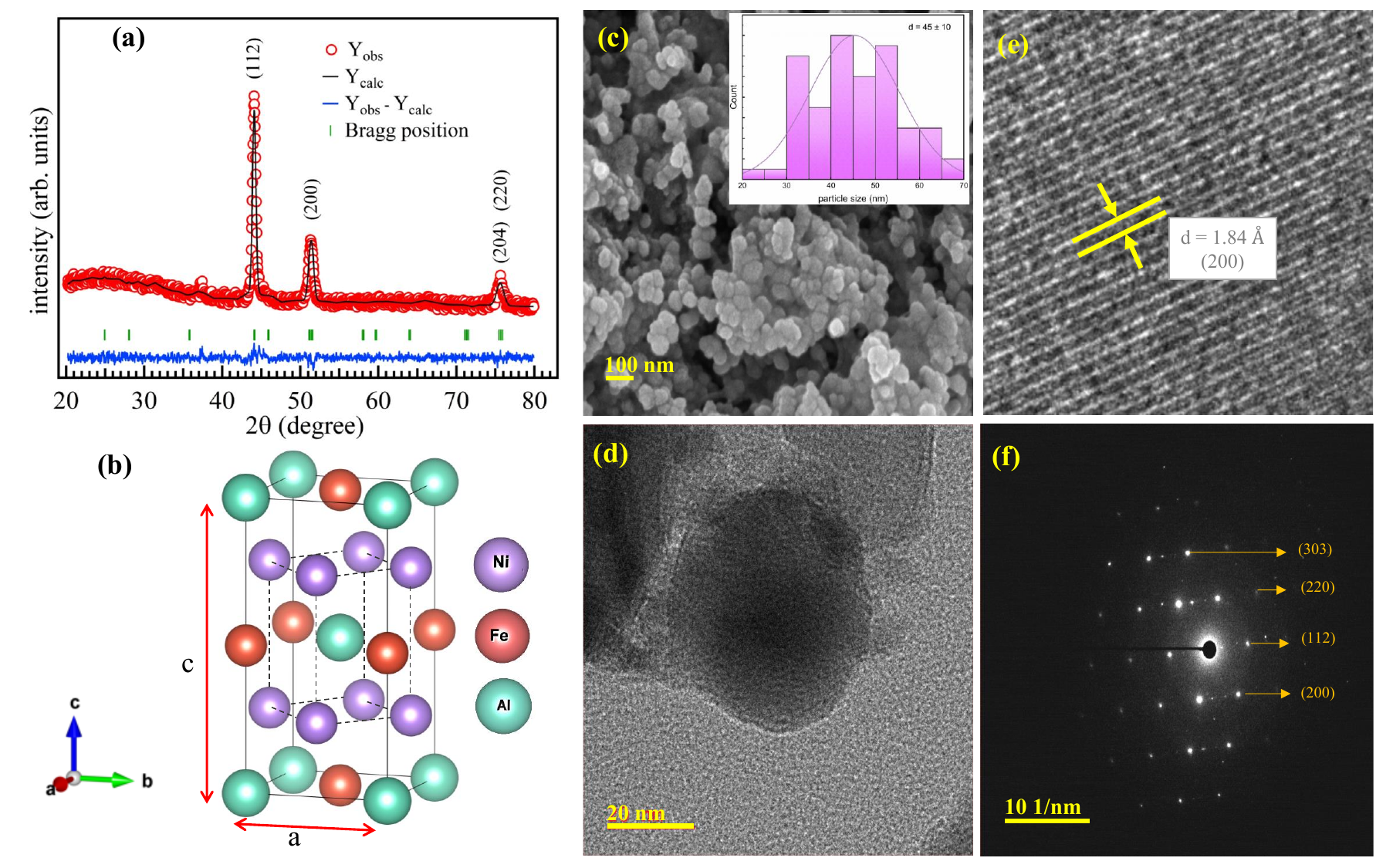}
\caption {(a) The XRD pattern for Ni$_2$FeAl NPs recorded at room temperature, with the observed data points, simulated curve, difference in observed and simulated data and Bragg positions, denoted by open red circles, solid black line, solid blue line, and vertical green markers, respectively, (b) the unit cell (space group I4/$mmm$) where $a=$ 3.556 \AA~and $c/a=1.42$, with Ni (violet), Fe (red), Al (cyan) atoms. (c) The FE-SEM micrographs along with distribution of particle size in the inset, (d-f) the HR-TEM image and the SAED pattern. }
\label{FESEM}
\end{figure*}

\section{Experimental and Computational Details}

Ni$_2$FeAl nanoparticles were synthesized following the procedure 
detailed in ref.~\cite{SI}. The crystallographic structure was studied using Malvern 
PANalytical conventional diffractometer with Cu K$_{\alpha}$ radiation 
($\lambda = 1.5406$\AA). Morphology and elemental composition were 
characterized using field-emission scanning electron microscope (FESEM), 
high resolution transmission electron microscope (HR-TEM) and energy-dispersive 
spectroscopy (EDS) measurements. Magnetic measurements were performed 
using MPMS3 SQUID magnetometer (Quantum Design, USA), and oven assembly 
was used for the high temperature measurements. Temperature-dependent electrical resistivity was measured using the standard four-probe technique in the physical property measurement system (PPMS) Evercool II (Quantum Design, USA) for the range of 2--300 K under both 0 and 10 kOe applied magnetic fields. 
Additionally, field-dependent longitudinal resistivity $\rho_{xx}$ was 
recorded in external magnetic fields up to 70 kOe, with the 
field oriented perpendicular to the sample plane and the electrical 
current applied along the sample’s length.

First-principles calculations were performed using density functional 
theory (DFT) as implemented in the {\em Vienna ab initio simulation 
package} (VASP) \cite{Kresse_PRB_96} employing a plane-wave pseudopotential 
approach. The projector augmented wave (PAW) method was used to accurately 
represent the valence electrons' wave functions near the atomic cores, 
and Perdew, Burke and Ernzerhof (PBE) functional within the 
generalized-gradient-approximation framework was applied for 
the exchange-correlation \cite{Kresse_PRB_99}. To determine the ground 
state configuration, the unit cell was optimized using full relaxation 
calculations with energy convergence criterion of 10$^{-8}$ eV and atomic 
force tolerance of 10$^{-7}$ eV/\AA. The self-consistent-field calculations
were carried out using a Monkhorst-Pack \cite{Monkhorst_PRB_76} {\it k} 
mesh grid of 8$\times$8$\times$8. A plane wave basis with an energy cut-off 
400 eV is used in all the calculations. The MCA energy was evaluated 
as the energy difference between two spin orientations, $E_{\rm MCA} = E_{(100)} - E_{(001)}$. The respective total energies have been calculated by initial collinear ground-state calculation, followed by non-self-consistent inclusion of spin-orbit coupling for different spin orientations. Phonon dispersion was subsequently calculated and examined using the finite displacement method as implemented in PHONOPY \cite{Togo_SM_2015} to check the structural stability of the studied system.

\section{Results and Discussion}

{\bf Crystal Structure and Magnetic Properties}: The XRD pattern measured at room temperature, shown in Fig.~\ref{FESEM}(a), confirms the single-phase crystallization of Ni$_2$FeAl NPs in tetragonal (space group I4/$mmm$, No.~\#139) structure. Rietveld refinement yields lattice parameters $a=$ 3.556 \AA~and $c/a=1.42$, Fig.~\ref{FESEM}(b). The quality of refinement is indicated by R$_P$ = 3.50, R$_{\rm exp}$ = 3.52 and $\chi^2$ = 1.17. The average crystallite size (D$_v$) is estimated to be 25 nm using the $\text{Sch\"{e}rrer}$ formula, D$_v$=$\frac{K\lambda}{\beta cos\theta}$, where {\it K}, $\lambda$ and $\beta$ are the dimensionless shape factor, wavelength of x-ray and full-width at half the maximum intensity of diffraction peak, respectively. Elemental analysis via FESEM-EDS measurements, shown in Fig.~S2 of \cite{SI}, confirm the stoichiometric composition of the synthesized nanoparticles.  The FE-SEM and HR-TEM micrographs (Fig. \ref{FESEM}(c,d)) reveal spherical nanoparticles with an average particle size of $\sim$45 nm and a moderate size dispersion, as determined using ImageJ software [see inset of Fig.~\ref{FESEM}(c)]. Complementary selected area electron diffraction (SAED) and fast Fourier Transform (FFT) analyses of the HR-TEM images (Figs.~\ref{FESEM}(e, f)) further verify the structural phase, with the interplanar spacing denoted in Fig.~\ref{FESEM}(e) aligning with the (200) plane also identified from XRD, hence, confirming the presence of an ordered I4/$mmm$ phase.

\begin{figure*}
\includegraphics[width=7.2in]{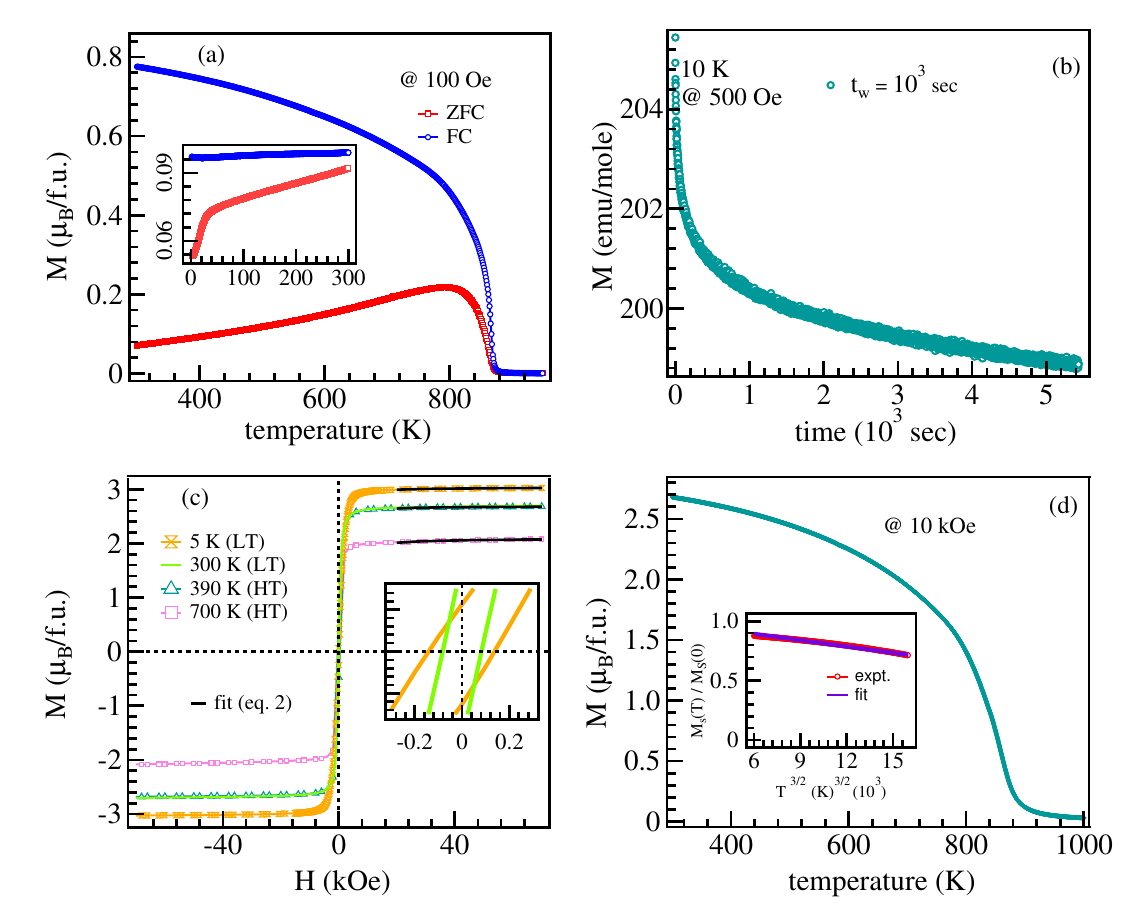}
\caption {The thermo-magnetization curve recorded across the temperature range 300-1000 K at 100 Oe applied magnetic field for ZFC (open red squares) and FC (open blue circles) protocols. The inset illustrates low temperature thermo-magnetization curves across 2-300~K, recorded at 100 Oe. (b) The time dependence of iso-thermal remanent magnetization M(t) measured at 10 K. (c) Isothermal magnetization measured at 5, 300~K in low temperature (LT) assembly and 390 and 700~K in high temperature (HT) assembly, with inset showing the zoomed view of isotherms at 5 and 300~K. The high field region of the isotherms are fitted using the law of approach to saturation. (d) Thermo-magnetization curve recorded across 300-1000~K under an applied magnetic field of 10 kOe. The inset presents the normalized curve plotted as a function of T$^{3/2}$ and fitted using Bloch's function.}
\label{MT-MH}
\end{figure*}

To investigate the magnetic behavior, temperature-dependent magnetization 
measurements are performed under zero-field-cooled (ZFC) and field-cooled (FC) 
protocols in an applied field of 100 Oe. As shown in inset of Fig.~\ref{MT-MH}(a), 
the ZFC (open red squares) magnetization gradually decreases down to 50~K, followed by a pronounced drop at lower temperatures. A similar reduction in the FC (open blue circles) magnetization, along with a substantial bifurcation between ZFC and FC curves, suggests the coexistence of AFM-FM interactions at low temperatures, leading to magnetic frustration. This behavior likely arises from spin-glass-like freezing and/or magnetic blocking effects associated with 
anisotropy and domain interactions within the FM matrix \cite{Read_JMMM_1984, Ajay_PRB_2024}, as discussed in detail later. At higher temperatures, as discernible from Fig.~\ref{MT-MH}(a), both ZFC and FC curves continue to exhibit significant bifurcation. The magnetic transition temperature is estimated to be T = 874~K from the first-order derivative of magnetic susceptibility, closely matching the Curie temperature ($\theta_{\rm C}=$ 877~K), obtained via Curie-Weiss fitting of the inverse magnetic susceptibility, see Fig.~S3 of \cite{SI}. 

The signature of magnetic frustration as hypothesized earlier is investigated through isothermal remanent magnetization (IRM) measurements, depicted in Fig.~\ref{MT-MH}(b). The sample was field-cooled to 10 K, and held for 1000 seconds before the applied field is removed and data are recorded. The remanent magnetization exhibits a clear temporal decay, persisting beyond one hour, indicating slow spin dynamics, characteristic of glassy or frustrated magnetic states. A detailed quantitative analysis, presented in the Fig.~S4 of \cite{SI}, suggests the coexistence of ferromagnetic clusters within a glassy matrix, likely arising from inter-particle interactions or localized magnetic ordering within the nanoparticle ensemble \cite{Gabay_PRL_1981,Anand_PRB_2012}. The low-temperature spin dynamics is further examined through {\it ac}-susceptibility measurements recorded at an excitation field of $H_{ac}=$ 2.5 Oe over a frequency range of 9-698 Hz. A pronounced cusp in $\chi_{ac}'$ and the corresponding peak in $\chi_{ac}''$ were observed, both shifting toward higher temperatures with increasing frequency, accompanied by reduction in amplitude, as shown in Fig.~S4(b, c) of \cite{SI}. This frequency dispersion reflects thermally activated spin dynamics and frequency-dependent domain reorganization. Detailed analysis presented in \cite{SI} confirms strong inter-cluster interactions, consistent with cooperative spin freezing in a cluster-glass regime. These observations collectively indicate that the magnetic relaxation arises from correlated nanoparticle ensembles, rather than from the independent thermally activated superparamagnetic relaxation of isolated spins \cite{Jonsson_ACL_2003}. 
  
\begin{figure*}
\includegraphics[width=7.4in]{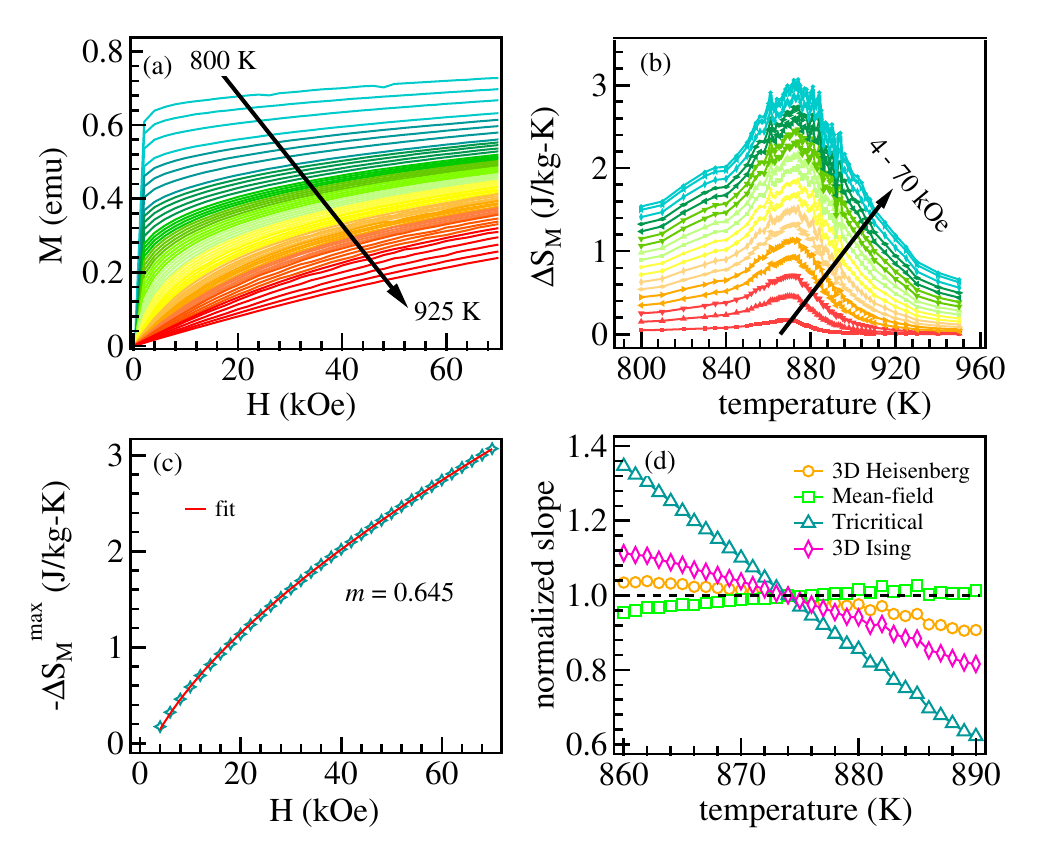}
\caption {(a) Isothermal magnetization curves for temperature range 800--925 K, 
	up to 70 kOe applied magnetic field. (b) The magnetic entropy change 
	$\Delta S_{\rm M}$ versus temperature plotted as a function of magnetic 
	field (4--70 kOe). (c) The variation of maximum 
	entropy change as a function of applied magnetic field for critical 
	isotherm ($T = T_{\rm C}$) (d) Normalized slopes (NS) for different 
	universality classes in the vicinity of $T_{\rm C}$ for Ni$_2$FeAl.}
\label{MCE}
\end{figure*}
  
 Isothermal magnetization curves measured at selected temperatures, shown in Fig.~\ref{MT-MH}(c), confirm the soft ferromagnetic nature of the system. At 5 K, the saturation magnetization value ($M\rm_s$) reaches 3.02 $\mu_B$/f.u., with a coercive field ($H_{\rm C}$) of 140 Oe, which at 300 K decreases to 80 Oe. When the particle becomes sufficiently small, the energy associated with forming multiple magnetic domains exceeds that of maintaining a single/mono-domain configuration. Therefore, the critical size ($D_{\rm cr}$) beyond which a multi-domain 
structure becomes energetically favorable is given by \cite{Kittel_RMP_1949}: 
\begin{equation} 
	D_{\rm cr}=\frac{72\sqrt{AK_V}}{4\pi M{\rm_s^2}},
  \label{Dcr}
\end{equation}
where $A$ is the exchange constant, $K_{\rm V}$ is the bulk anisotropy constant ($K_{\rm eff}$ in this case) and $M_{\rm s}$ is the saturation magnetization. To estimate the $D_{\rm cr}$, we first evaluate the magnetic anisotropy using the law of approach to saturation (LAS) \cite{Deka_JMMM_2016}. In the high-field regime, the magnetization obeys the relation
\begin{equation} 
	M(H) =  M_{\rm s} \left(1-\frac{c}{H^2}\right).
  \label{LAS}
\end{equation}
Here, the constant $c$ is related to the effective magnetic 
anisotropy constant $K_{\rm eff}$ as
\begin{equation} 
	K_{\rm eff}= \mu_0 M_{\rm s} \left(\frac{105c}{8}\right)^\frac{1}{2},
\label{K}
\end{equation}
where $\mu_0$ is the permeability of free space. Fitting equation \eqref{LAS} 
to the high-field magnetization data in Fig.~\ref{MT-MH}(c), we find 
$K_{\rm eff}$ to be 0.238 MJ/m$^3$ and 0.216 MJ/m$^3$ at 5~K and 
390~K, respectively. The observed decrease in anisotropy with increasing temperatures is 
consistent with the trends reported for other magnetic NPs \cite{Ahmad_JMMM_2019, Srivastava_PCCP_2024}. Such high anisotropy constant in NPs could be attributed to several factors including, inter-particle interactions due to agglomeration \cite{Cardona_ApplSci_2019}, enhanced surface-to-volume ratio leading to broken exchange bonds at the particle surface and, strong dipolar interaction in the single-domain regime \cite{Paswan_JPCS_2021}.

Furthermore, the exchange constant $A$ and spin-wave stiffness constant 
$D$ are key parameters characterizing the strength of exchange 
interactions \cite{Trudel_JPDAP_2010}. These parameters are related 
as \cite{Kubota_JAP_2009}
\begin{equation} 
  A(T)=\frac{M_{\rm s}(T) D(T)}{2g\mu_B},
  \label{Aeq}
\end{equation}
where $g$ is the Lande $g$-factor and $M_{\rm s}$ is the saturation magnetization. Both $A$ and $D$ can be extracted through temperature-dependent magnetization measurements \cite{Ritchie_PRB_2003}. Here, the thermal decay of $M_{\rm s}$ arises from the thermal excitation of spin waves, 
and is described by the Bloch's law \cite{Kittel} as
\begin{equation} 
	\frac{M_{\rm s}(T)}{M_{\rm s}(0)} = 1 - BT^{3/2},
  \label{Blocheq}
\end{equation} 
where $M_{\rm s}(0)$ is the saturation magnetization at 0~K and 
$B$ is the Bloch coefficient, expressed as 
\begin{equation} 
	B = 2.612\frac{g\mu_B}{M_{\rm s}(0)}\left(\frac{k_B}{4\pi D}\right)^{3/2}
	\label{Beq}
\end{equation}
Fitting the magnetization data [inset of Fig.~\ref{MT-MH}(c)] with 
equation (\ref{Blocheq}), and using equations (\ref{Beq}) and (\ref{Aeq}), we 
extracted $A=2.296\times$10$^{-6}$ erg/cm 
and $D = 8.56$ meV(nm)$^2$. These values of $A$ and $D$ are consistent with the values reported for Heusler alloys, such as Co$_2$MnSi ($A = 2.35 \times10^{-6}$ erg/cm, $D = 5.86$ meV(nm)$^2$) \cite{Kubota_JAP_2009}, Co$_2$MnAl ($A = 0.48\times10^{-6}$ erg/cm, $D = 1.90$ meV(nm)$^2$) \cite{Kubota_JAP_2009} and Co$_2$MnGe ($A = 0.98\times10^{-6}$ erg/cm) \cite{Gaier_JAP_2008}. 

In Heusler alloys, the total magnetic moment and Curie temperature ($T_{\rm C}$) are often correlated with the valence electron count ($N_{\rm v}$). Interestingly, the spin-wave stiffness constant $D$ also scales with $N_{\rm v}$ for a given crystallographic ordering \cite{Kubota_JAP_2009}. Consequently, compounds with higher $N_{\rm v}$ exhibit reduced temperature sensitivity of spin polarization, assuming minimal contribution from other spin depolarizing mechanisms \cite{Kubler_PRB_2007}. This behavior has also been experimentally validated in Co$_2$FeSi based tunnel junctions, where tunneling magnetoresistance decays slowly with increasing temperature due to high values of $D$ and $M_{\rm s}$ \cite{Oogane_JAP_2009}. In this context, substantial values of $A$, $D$ and $K_{\rm eff}$ observed for Ni$_2$FeAl NPs in our study highlight their potential for spintronic applications. 

Using the extracted values of $A$ (2.296$\times$10$^{-6}$ erg/cm), $D$(8.56 meV(nm)$^2$) and K$_{eff}$(2.379$\times$10$^5$ Jm$^{-3}$) in equation (\ref{Dcr}), we calculate the critical single-domain size $D_{\rm cr}$ for Ni$_2$FeAl NPs, which found to be 73 nm. Since the average diameter is 45 nm, i.e. $d < D_{\rm cr}$, the synthesized particles can be classified as single-domain. As mentioned earlier, magnetic nanoparticles offer a compelling alternative to conventional bulk magnetocaloric materials due to 
their ease of integration into thin-film architectures, tunable entropy change 
via particle size modulation, and enhanced thermal exchange efficiency stemming 
from their high surface-to-volume ratio \cite{Poddar_JMMM_2006}. Interestingly, 
when the particle size is reduced to the single-domain regime, the magnetic 
entropy change is significantly enhanced compared to their bulk 
counterparts \cite{Michael_JMMM_1992}. Next we investigate 
the magnetocaloric response in Ni$_2$FeAl. For this purpose, the isothermal magnetization measurements are performed near $T_{\rm C}$ (874~K) with a temperature step of $\Delta T = 1$~K and applied magnetic fields up to 70 kOe, see Fig.~\ref{MCE}(a). The isothermal entropy change ($\Delta S_{\rm M}$) is calculated using Maxwell's equation \cite{PriyankaPRB, PriyankaJAP}, 
\begin{equation} 
   \Delta S_{\rm M} =\mu_0 \int_{0}^{H_{max}} 
	\left(\frac{\partial{M(T,H)}}{\partial{T}}\right)_H dH
   \label{entropy}
\end{equation}
The resulting $\Delta S_{\rm M}$ values for fields ranging from 
4 to 70 kOe are shown in Fig.~\ref{MCE}(b) where we find a well-defined peak centered at $T_{\rm pk}$, with a slight shift towards higher 
temperatures as the applied field increases. Furthermore, the magnitude of $\Delta S_{\rm M}$ grows with increasing field strength and the maximum $\Delta S_{\rm M}$ reaches 3.1 J.kg$^{-1}$K$^{-1}$ at 70 kOe, demonstrating a substantial magnetocaloric  effect (MCE) that is comparable with other nanoparticulate 
Heusler systems \cite{Ahmad_JMMM_2021}.

Further, we examine the nature of magnetic phase transition utilizing the Banerjee criterion, where a second-order phase transition (SOPT) is characterized 
by Arrott curves ($M^2 {\rm vs.} H/M$) exhibiting positive slope across all 
isotherms measured above and below $T_{\rm C}$ \cite{Banerjee_PL_1964}. 
In Fig.~S6(c) \cite{SI}, we observe positive slope for all the measured 
isotherms, confirming the second-order phase transition in the present sample. The SOPT character is further substantiated using the universal scaling approach, as discussed in detail in \cite{SI}, see Fig.~S5. Additionally, the magnetic entropy change corresponding to the critical isotherm ($T = T_{\rm C}$) is plotted as a function of applied field using the relation $\Delta S_{\rm M}^{max}$ $\propto$ $H^{m}$, see Fig.~\ref{MCE}(c). The extracted value of $m=0.645$ is in good 
agreement with the mean-field model prediction ($m=0.66$). Consequently, 
modified Arrott plots [$M^{1/\beta} {\rm vs.} (H/M)^{1/\gamma}$] corresponding
to critical exponents predicted for the tricritical ($\beta=0.25$, $\gamma=1$), 
3D Heisenberg ($\beta=0.365$, $\gamma=1.386$), mean-field ($\beta=0.5$, $\gamma=1$), and 3D Ising ($\beta=0.325$, $\gamma=1.24$) models \cite{Kaul_JMMM_85}, are constructed in Figs.~S6(a--d) of \cite{SI}. 
To determine which model produces the best fit, in Fig.~\ref{MCE}(d), we 
show the normalized slope ($=S(T)/S(T_{\rm C})$), where the slope at 
any given temperature is normalized to the slope at $T_{\rm C} = 874$ K. We 
find a least deviation from unity for the mean-field model, suggesting 
it successfully characterizes the magnetic interactions in the studied Ni$_2$FeAl nanoparticles.

\begin{figure}
\includegraphics[width=3.4in]{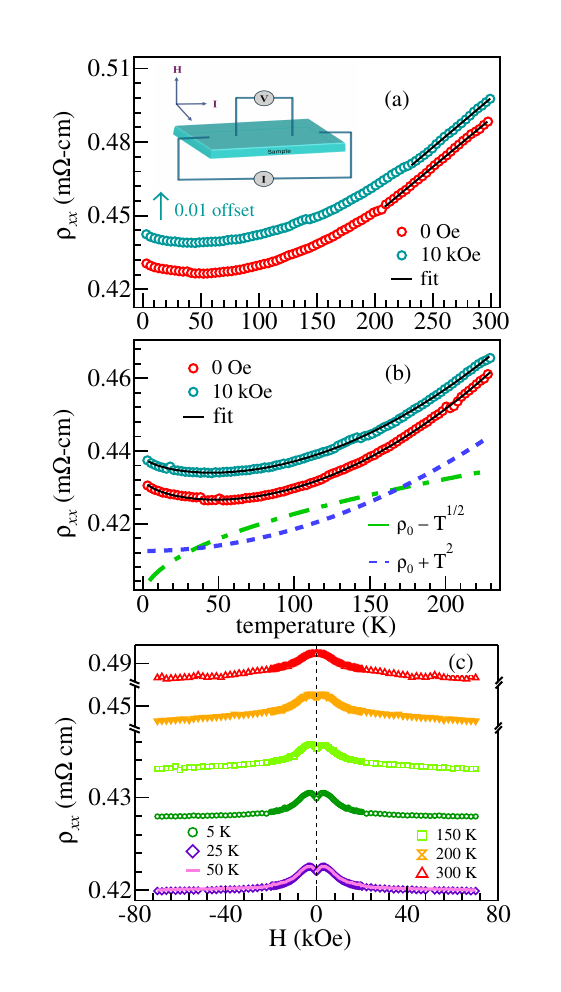}
\caption {(a) The temperature dependence (3--300 K) of longitudinal $\rho_{xx}$ under different magnetic field. The black solid curves represent the fit corresponding to linear temperature dependence. (b) the $\rho_{xx}$ versus $T$ (3--230 K), fitted using $T^{1/2}$+$T^2$ relation (solid black). The dashed curves represent the best fit corresponding to $T^{1/2}$ (green) and $T^2$ (blue) dependence. (c) The temperature dependent resistivity curves under applied field in the range of -70 to 70 kOe.}
\label{RT}
\end{figure}

{\bf Transport properties}: The magnetotransport behavior of Ni$_2$FeAl is investigated via longitudinal resistivity ($\rho_{xx}$) measurements under the applied magnetic fields of 0 and 10 kOe, as shown in Fig.~\ref{RT}(a). 
At high-temperatures, the $\rho_{xx}$ exhibits monotonic decrease with decreasing 
temperature, primarily governed by the suppression of phonon-mediated 
scattering processes \cite{Tong_PRB_2020}. A linear fit ($\rho \propto T$) yields excellent agreement in the ranges of 300--209~K (0 Oe) and 300--226 K (10 kOe), consistent with the electron-phonon scattering. Deviations from linearity below 216 K (0 Oe) and 230 K (10 kOe), indicate a transition in the predominant scattering mechanism. In the intermediate temperature range, the $\rho_{xx}$ best scales with $\rho_{xx}\propto T^2$ dependence, typically attributed to the electron-magnon scattering \cite{Saha_PRB_2023}. However, the electron-electron scattering can also lead to $T^2$ scaling \cite{Saha_PRB_2023}. To distinguish between these mechanisms, typically an external magnetic field is applied, which induces a gap in the spin-wave spectrum. This gap suppresses the thermally excited magnons, thereby inhibiting electron-magnon scattering, whereas electron-electron Coulomb scattering remains largely unaffected \cite{Madduri_PRB_2017}. The comparable slope of the $T^2$ fits under 0 and 10 kOe fields (see Fig.~S7 of \cite{SI}), suggest that electron-electron scattering is the dominant contribution in this temperature range. This is further supported by the large spin-wave stiffness constant $D$ extracted from the magnetic measurements, which implies a diminished magnon population and correspondingly weak magnon-mediated scattering \cite{Trudel_JPDAP_2010}. At low temperatures, both the $\rho_{xx}$ curves exhibit a minimum near $T\sim45~$K, followed by an anomalous upturn upon further cooling. This anomalous behavior is indicative of competing conduction mechanisms. Several possibilities were examined, including spin-polarized tunneling across grain boundaries \cite{Helman_PRL_1976}, Kondo-type scattering from spin disorder \cite{Kondo_PTP_1964}, and quantum corrections to conductivity (QCC), encompassing weak localizations (WL) \cite{Bergmann_PR_1984} and electron-electron 
interaction (EEI) effects \cite{Lee_RMP_1985}. The spin-polarized tunneling 
mechanism involves electron transfer between adjacent grains with misaligned 
magnetic moments, wherein the tunneling probability is modulated by exchange 
energy. This phenomenon has been observed in granular systems such as Ni 
films \cite{Helman_PRL_1976}. Conversely, the Kondo effect arises from 
conduction electron scattering off of localized magnetic moments or spin 
clusters, resulting in a logarithmic temperature dependence of the resistivity, 
$\rho(T) = - C_{\rm K} {\rm ln}(T)$ \cite{Kondo_PTP_1964}. Such behavior is reported in granular alloys like Au-Ni and Co-Cu, where low-temperature 
resistivity is predominantly governed by Kondo scattering from spin clusters. 
However, application of magnetic field suppresses these scattering events and 
diminishes the associated resistivity upturn \cite{Dhara_PRB_2016, Fabietti_PRB_2010, Mei_PRB_1986}. 

In the present sample, neither the temperature at which the resistivity 
minimum occurs ($T_{\rm min}$) nor the depth of the minimum, exhibit significant 
variation under the applied magnetic field. This effectively rules 
out spin-polarized tunneling and Kondo-type scattering as the dominant 
mechanisms for the observed low-temperature resistivity upturn. Instead, 
quantum corrections to conductivity (QCC), particularly those arising from 
disorder-enhanced electron-electron interactions (EEI), provide a more 
plausible explanation. Within Fermi-liquid theory, resistivity in a weakly 
disordered system typically scales as $\rho \propto $ A$T^2$ when EEI dominate. However, in the systems with moderate to strong disorder, EEI corrections lead to a characteristic $\rho_{xx} \propto T^{1/2}$ dependence \cite{Lee_RMP_1985}. The total low-temperature resistivity, incorporating QCC effects can be modeled as: 
\begin{equation} 
  \rho = \rho_0-|C_{\rm EEI}| T^\frac{1}{2}+|C_{\rm e}|T^2,
  \label{RT-low}
\end{equation}
where $\rho_0$ is the residual resistivity. The clear $T^{1/2}$ dependence, coupled with the field-insensitivity of the resistivity minimum, strongly supports the dominance of disorder-enhanced EEI in governing low-temperature transport behavior \cite{Muthuselvam_JALCOM_2012}. 
 
 \begin{figure*}
  \includegraphics[width=6.8in]{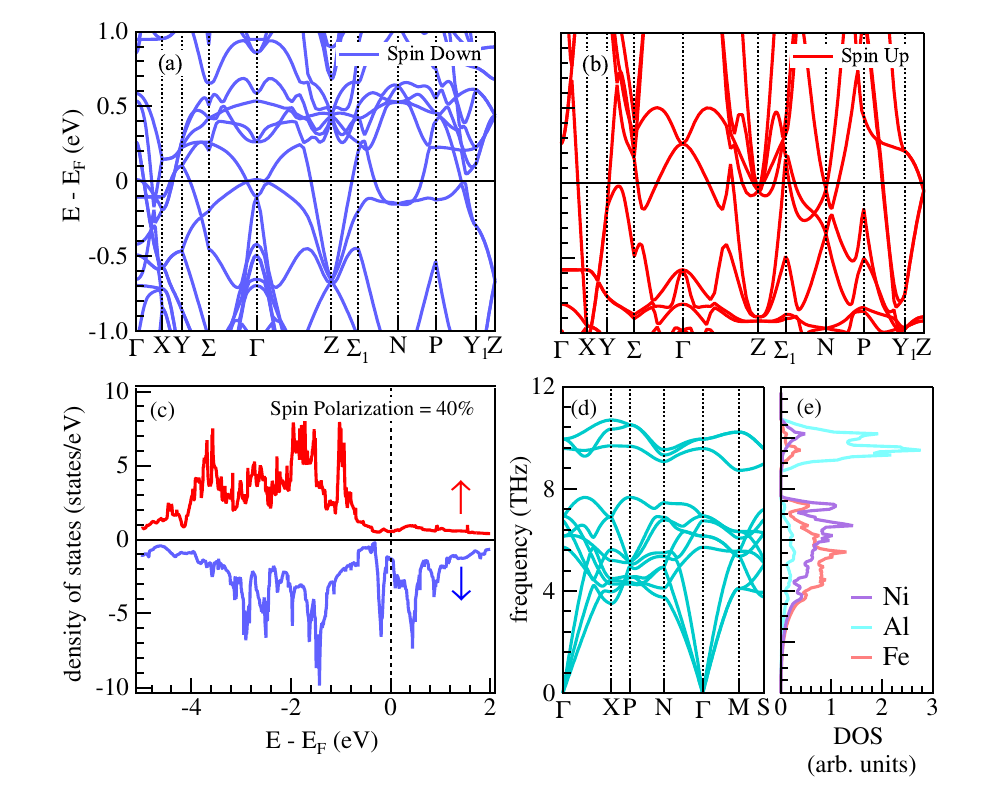}
  \caption {(a, b) The electronic band structure near the Fermi level for minority 
	and majority spin states, respectively. (c) The spin polarized total 
	density of states of Ni$_2$FeAl for near Fermi level region. 
	(d) The phonon dispersion and (e) projected phonon density of states 
	for Ni, Fe and Al atoms for I4/$mmm$ phase.}
	\label{DOS}
\end{figure*}
 
Additional support for disorder-driven conduction comes from the low 
residual resistivity ratio (RRR), defined as $\rho(300\rm K)/\rho(45\rm K)$, 
which is found to be 1.15 for 0~Oe, consistent with moderately disordered systems \cite{Tong_PRB_2020, Young_JAP_2014}. To further investigate the field dependence of resistivity, 
the magnetoresistance (MR) measurements are performed, as shown in Fig.~\ref{RT}(c). The resistivity curves exhibit an upturn at low fields ($H\sim$ 5~kOe), followed by a sign-reversal and eventual 
saturation at higher fields. Typically in ferromagnets, the MR reflects the interplay of ordinary cyclotron motion and spin-dependent scattering, where the former contributes positively while latter yields a quasi-linear negative MR above the saturation field. The observed low-field dip hence originates from the steep increase in magnetization prior to saturation, while the downturn or sign reversal near saturation field arises from the dominant spin-scattering  \cite{Tong_PRB_2020, Saha_PRB_2023}. 

\begin{figure*}
  \includegraphics[width=\textwidth]{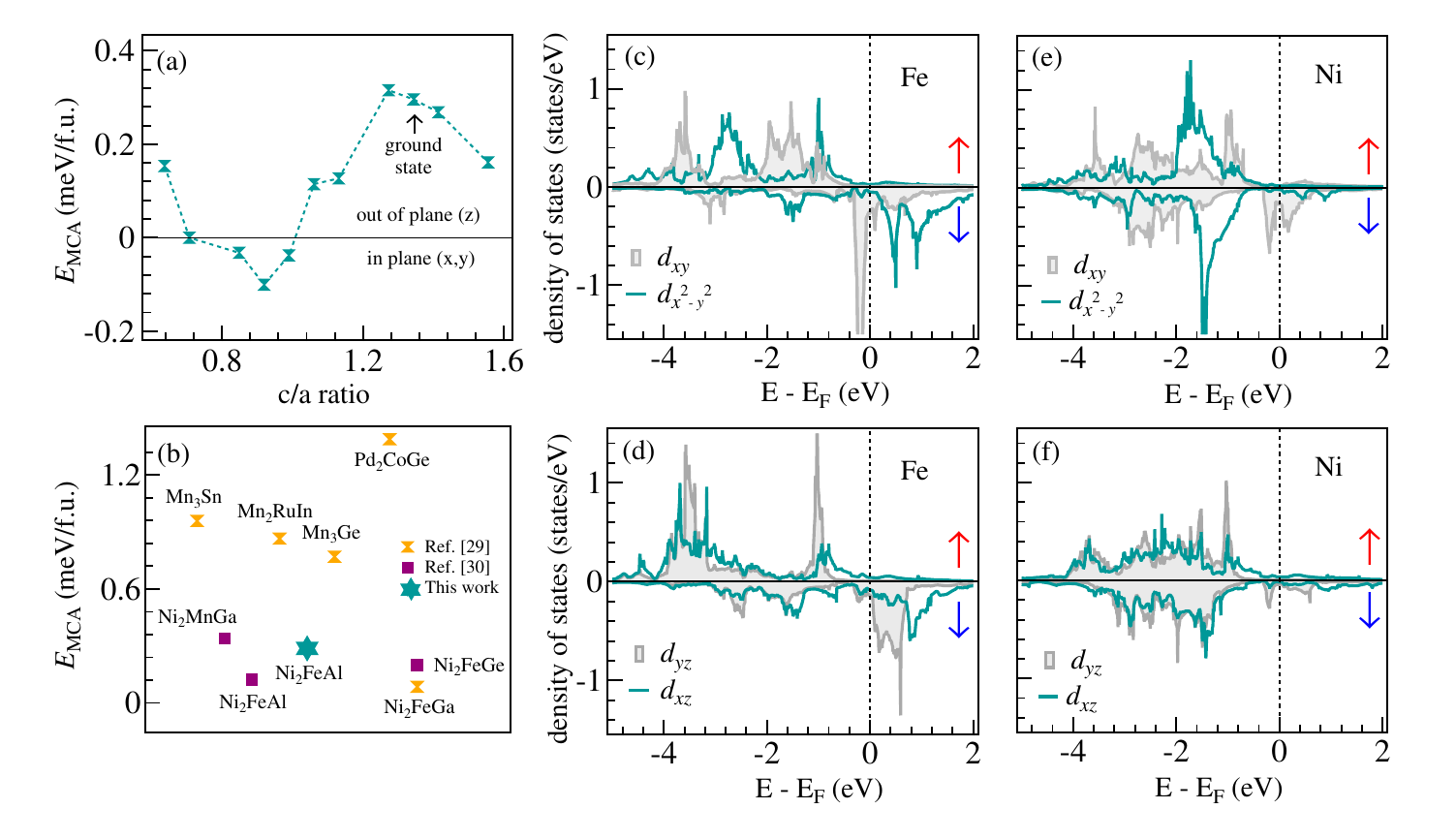}
  \caption {Dependence of magnetocrystalline anisotropy on the 
	lattice ratio $c/a$ for Ni$_2$FeAl Heusler alloy. (b) Theoretically reported $E_{\rm MCA}$ values of Heusler alloys and related systems. Calculated orbital resolved density of states of Ni$_2$FeAl, 
	the middle panel shows PDOS for Fe-$d$ states (c) ($d_{xy}$, 
	$d_{x^2-y^2}$) and (d) ($d_{xz}$, $d_{yz}$) pair, whereas 
	the Ni-$d$ states (e) ($d_{xy}$, $d_{x^2-y^2}$) and (f) ($d_{xz}$, $d_{yz}$) are given in the right panel.}
  \label{MAE}
\end{figure*}

{\bf First-principles Simulations}: Moreover, we performed first-principles 
calculations based on density functional theory to explore the electronic 
structure, dynamical stability and magnetic properties of Ni$_2$FeAl and to substantiate our experimental observations. While earlier theoretical studies have considered both the cubic (L2$_1$) and tetragonal (I4/$mmm$) phases for this sample \cite{Wen_I_2018, Sehoul_JSNM_2025}, 
comparative analyses indicate that I4/$mmm$ configuration is energetically 
favored in Ni-based Heusler alloys \cite{Benhizia_CCM_20, Wen_I_2018}. 
Notably, Wen {\it et al.} identified two local energy minima for $c/a > 1$ 
and $c/a < 1$ in Ni$_2$FeAl, reinforcing the relative stability of the 
tetragonal structure over the cubic phase \cite{Wen_I_2018}.
The structural optimization calculations are executed using a 2$\times$2$\times$2 
supercell, which confirm that the system stabilizes in I4/$mmm$ phase, with lattice parameters $a=$ 7.355 \AA~and $c/a=$ 1.34. The calculated total magnetic moment is found to be 3.22 $\mu_{\rm B}$/f.u., 
which slightly exceeds the experimental value of 3.02 $\mu_{\rm B}$/f.u. 
at 5~K, see Fig.~\ref{MT-MH}(b). Furthermore, the total and local magnetic 
moments are consistent with the previously reported values, as summarized 
in Table \ref{Tlp}. The magnetic moment is primarily localized on the Fe 
atoms 2.56 $\mu_{\rm B}$/f.u., followed by Ni atoms with 0.34 $\mu_{\rm B}$/f.u., 
while Al atoms exhibit a weak antiparallel moment of 
approximately -0.03 $\mu_{\rm B}$/f.u.

Now, we compute spin-resolved density of 
states (DOSs) and electronic band structure for Ni$_2$FeAl, as shown in Fig.~\ref{DOS}. The electronic band structures for both spin channels, panels (a) and (b), 
confirm the metallic nature of the sample. The total DOS, Fig.~\ref{DOS}(c), shows 
finite states at the Fermi level for both the spin channels, and the 
calculated spin polarization is around 40\%. Our calculations are consistent with the 
previous values reported for I4/$mmm$ (35\%) and L2$_1$ (46\%) phases \cite{Qawasmeh_JAP_2012}. 
The partial DOS, presented in Fig.~S8 of \cite{SI}, reveal that states 
near the Fermi level are primarily composed of Ni-$3d$ and Fe-$3d$ orbitals, 
with a strong hybridization between them playing a central role in determining 
the magnetic and transport characteristics. Further, to assess 
the dynamical stability of Ni$_2$FeAl, we examine phonon dispersion using {\em ab-initio} finite displacement method. As can be observed from Fig.~\ref{DOS}(d), the phonon spectra exhibit no imaginary modes, confirming the dynamical stability of I4/$mmm$ phase. The projected phonon density of states, Fig.~\ref{DOS}(e), 
show that Ni and Fe dominate the low-frequency acoustic modes, while Al 
participates primarily in higher frequency optical modes. 

\begin{figure*}
  \includegraphics[width=6.75in]{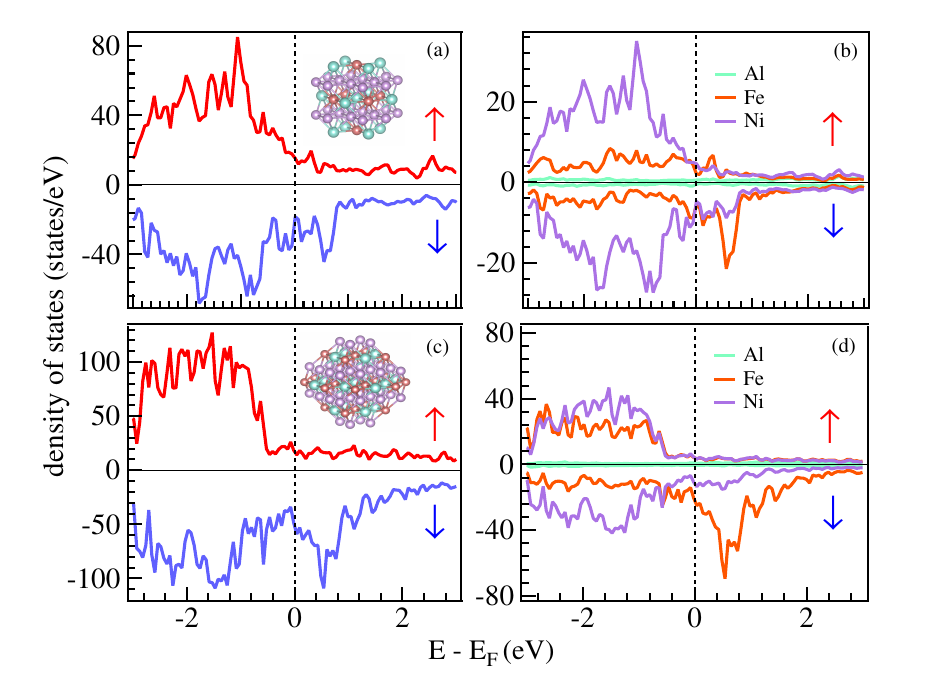}
  \caption {The total and atom-resolved spin polarized DOSs of (a, b) NC$_{43}$, Ni$_{24}$Fe$_{6}$Al$_{13}$ and (c, d) NC$_{79}$, Ni$_{40}$Fe$_{22}$Al$_{17}$ systems.}
  \label{n-dos}
\end{figure*}

{\bf Magnetocrystalline anisotropy}: Having established the structural stability, we now turn to the magnetic anisotropy, which originates from tetragonal lattice distortion that induces a preferred magnetization axis. Total energy calculations with spin-orbit coupling (Table~I of \cite{SI}) reveal that the [001] direction is energetically favored over [100]. The resulting magnetocrystalline anisotropy energy ($E_{\rm MCA}$) is computed as the total energy difference between magnetization aligned along the [100] and [001] directions, yielding 0.287 meV/f.u. (0.987 MJ/m$^3$). After accounting for shape anisotropy, $E_{\rm sh} = \mu_0 M^2_{\rm s}V/2$ per unit cell of volume V, the volume magnetic anisotropy energy ($E_v$) is estimated using $E_v = E_{\rm MCA}-E_{\rm sh}$. We obtain $E_v=0.728$ MJ/m$^3$, which is of the same order as the experimentally extracted effective anisotropy constant $K_{\rm eff}=0.238$ MJ/m$^3$ at 5~K, derived via the 
LAS method. The quantitative agreement is noteworthy, given that the DFT results 
pertain to an ideal, defect-free system at 0~K, whereas the experimental anisotropy is derived indirectly from magnetization measurements for disordered system. The computed anisotropy also aligns well 
with previous theoretical estimates, for instance, 0.122 meV/f.u. for Ni$_2$FeAl reported by Herper {\em et al.} \cite{Herper_PRB_2018}. Discrepancies in the absolute magnitude of the anisotropy across different  theoretical studies primarily arise from differences in computational parameters, exchange–correlation functionals and methodological approximations \cite{Herper_PRB_2018,Parkin_PRM_2017}. Nevertheless, both the order of magnitude and sign of anisotropy remain robust, allowing meaningful trends to be established within consistent theoretical framework. We also compute the variation of $E_{\rm MCA}$ with tetragonal distortion ratio $c/a$, see Fig.~\ref{MAE}(a), which shows that the uniaxial (out-of-plane) anisotropy persists across most distortion values, 
except near the cubic limit. Further, we evaluate the anisotropy using Bruno's model, which relates the magnetic anisotropy energy (MAE) to the anisotropy of orbital magnetic moments \cite{Bruno_PRB_1989}. This approach successfully reproduces the orientation of easy-axis across the full $c/a$ range (see Table~II of \cite{SI}). Although, the calculated MAE for the Ni$_2$FeAl (0.287 meV/f.u.) is modest compared to $5d$-transition metals or rare-earth based systems, it is comparable to established anisotropic materials like Ni$_2$MnGa (0.338 meV/f.u.) \cite{Herper_PRB_2018} and other tetragonal Heusler systems \cite{Winterlik_AM_2013, Parkin_PRM_2017, Herper_PRB_2018}, shown in Fig.~\ref{MAE}(b).

However, it must be emphasized that magnetic anisotropy is not governed solely by 
lattice distortions. It also depends on orbital hybridization, band filling, 
and the underlying electronic structure \cite{Wollmann_PRB_2015, Umetsu_APL_2006}. On a microscopic level, SOC-mediated interactions between occupied and unoccupied electronic states near the Fermi level, determine the anisotropy strength. In particular, the rotationally degenerate ($d_{xz}$, $d_{yz}$) 
and ($d_{xy}$, $d_{x^2-y^2}$) orbital pairs are crucial, such that, unequal occupation within these pairs, i.e., pair-polarization favors uniaxial magnetization (along {\it z}-axis) when the highest occupied and lowest 
unoccupied states belong to the same orbital pair ($|\Delta L_z = 0|$ interaction) \cite{Herper_PRB_2018}. The SOC contribution to MAE can be quantified using the second-order perturbation theory \cite{vleck_PRB_1937}, as
\begin{equation} 
	\Delta E_{\rm LS} = \xi^2 \frac{|\langle k|\mathbf{L \cdot S}|n\rangle|^2}{E_k-E_n},
  \label{SOC}
\end{equation}
where $n$ and $k$ represent the unoccupied and occupied electronic states, 
respectively, and $\xi$ is the SOC constant. The matrix element in the numerator 
governs the directional dependence, while the denominator with smaller energy 
separation leads to enhanced MAE contribution \cite{Wollmann_PRB_2015, Bruno_PRB_1989}. 
To probe the orbital origin of anisotropy, we analyze the PDOS of Fe and Ni 
atoms, shown in Figs.~\ref{MAE}(c--f). The PDOS reveal clear signs of 
pair-polarization within both the degenerate pairs, indicating a preference 
for uniaxial anisotropy. The extracted pair-polarization indices, ($\Delta p$), 
yield $\Delta p = 0.30$ for ($d_{xy}$, $d_{x^2-y^2}$) and $\Delta p = 0.08$ for ($d_{xz}$, $d_{yz}$) in Fe. In contrast, Ni exhibits lower asymmetry, with $\Delta p = 0.04$ and $\Delta p=0.01$ for respective orbital pairs, 
as summarized in Table~III of \cite{SI}. These results suggest that Fe sublattice plays dominating role in governing the observed uniaxial magnetization in Ni$_2$FeAl. 
\begin{table}
  \caption{Total magnetic moment (in $\mu_{\rm B}$) and atom 
	resolved magnetic moments per formula unit.}
\begin{tabular}{lcccc}
  \hline
  \hline
  System & $m^{\rm Ni}$ & $m^{\rm Fe}$ & $m^{\rm Al}$ & $m^{\rm Total}$ \\
  \hline
  Ni$_{2}$FeAl (I4/$mmm$) (present) & 0.34 & 2.56 & -0.03 & 3.22   \\ 
  Ni$_2$FeAl (I4/$mmm$) \cite{Benhizia_CCM_20}& 0.38 & 2.6 & -0.04 & 3.23  \\
  Ni$_{2}$FeAl (L2$_1$) \cite{Benhizia_CCM_20}& 0.27 & 2.8 & -0.02 & 3.23  \\
  Ni$_{2}$FeAl (L2$_1$) \cite{Wen_I_2018}& - & - & - & 3.38 \\ 
  Ni$_{2}$FeAl (L2$_1$) \cite{Sehoul_JSNM_2025}& 0.26 & 2.79 & -0.02 & 3.22  \\ 
  Ni$_{2}$FeAl (L2$_1$) \cite{Qawasmeh_JAP_2012}& 0.24 & 2.57 & -0.02 & 2.99  \\  
  \hline
  \hline
\end{tabular}
\label{Tlp}
\end{table}
The pursuit of materials exhibiting substantial PMA and high tunneling magnetoresistance, either through 
strong spin polarization or Brillouin-zone filtering mechanisms \cite{Parkin_PRM_2017}, 
is pivotal for next-generation spin-transfer torque magnetic random-access 
memory (STT-MRAM) applications \cite{Parkin_PRM_2017}. In this context, 
the moderate yet robust anisotropy and favorable spintronic characteristics 
of Ni$_2$FeAl highlight its potential as a viable candidate for 
next-generation STT-based devices. 

{\bf Nanoclusters}: While the first-principles calculations on bulk samples confirm that Ni$_2$FeAl stabilizes in tetragonal structure with robust ferromagnetism and metallic behavior in both 
the spin channels, theoretical understanding of its nanoscale properties is vital to compare with experimental observations. To address this point, we employed first-principles 
simulations to examine the electronic and magnetic properties of bulk-like Ni$_2$FeAl 
nanoclusters (NCs), with particular focus on finite size and surface-driven effects. 
Two NCs consisting of 43 (NC$_{43}$: Ni$_{24}$Fe$_6$Al$_{13}$) and 79 atoms (NC$_{79}$: Ni$_{40}$Fe$_{22}$Al$_{17}$) are constructed and the structural 
relaxation reveal pronounced surface reconstruction, especially in NC$_{43}$, 
where near-surface atomic rearrangements dominate the local environment. 
Consequently, the magnetic and electronic properties of NC$_{43}$ are expected 
to be dominated by surface effects, while the larger NC$_{79}$ retains a more 
bulk-like core. Both clusters exhibit clear ferromagnetic order and total magnetic moment increases with cluster size; for example, 0.396 $\mu_{\rm B}$/atom for NC$_{43}$ and 0.966 $\mu_{\rm B}$/atom for NC$_{79}$, whereas the experimental value is found to be 0.755 $\mu_{\rm B}$/atom. While $p-d$ hybridization typically suppresses magnetic moments in Heusler 
alloys, the reduced hybridization at the surface enhances the local 
magnetic moments \cite{Kubler_PRB_1983} and Al atoms contribute 
minimally -0.028 and -0.038 $\mu_{\rm B}$/atom for NC$_{79}$ and NC$_{43}$, respectively, while Fe and Ni atoms primarily govern the magnetic response. 

The electronic structure (see Fig.~\ref{n-dos}) reveal the emergence 
of surface-derived states near the Fermi level and a pronounced size dependence in 
spin polarization. Notably, the spin polarization increases from 9\% in NC$_{43}$ to 44\% in the NC$_{79}$, approaching the bulk value reported for Ni$_2$FeAl (40\%). Our computed PDOS indicates 
that Ni-$3d$ states dominate the valence band, while Fe-$3d$ states contribute significantly to minority-spin conduction band, Figs.~\ref{n-dos}(b, d). In the NC$_{79}$, a notable enhancement in Fe minority-spin states is observed, correlating with an increased magnetic moment, potentially due to Fe-rich 
surface regions. These findings align with the earlier reports 
on Ni$_2$MnGa and Co$_2$MnGa clusters by Zayak {\it et al.} \cite{Zayak_JAP_2008, Zayak_PRB_2008}, which highlighted surface reconstruction as a key driver of electronic 
and magnetic evolution at reduced dimensions. Our results emphasize 
that surface-mediated modifications offer a strategic route to engineer spintronic functionality, particularly by spin polarization and local moment enhancement in Heusler alloys at nanoscale.

\section{Conclusion}

This study presents the synthesis of Ni$_2$FeAl nanoparticles via a template-free 
chemical route, complemented by {\it ab initio} first-principles calculations. 
The nanoparticles crystallize in I4/$mmm$ phase with an average 
particle size of 45 nm, exhibiting remarkable magnetic properties, like high 
$M_{\rm s}$ of 3.02 $\mu_{\rm B}$/f.u., significantly high magnetic anisotropy 
of 0.238 MJ/m$^3$ and low $H_{\rm C}$ of 140 Oe at 5~K. 
Magnetocaloric analysis reveals a magnetic entropy change 
of 3.1 J.kg$^{-1}$K$^{-1}$ at 70 kOe, highlighting the 
potential for energy-efficient cooling applications. Transport measurements 
indicate a low-temperature resistivity upturn, dominated by enhanced 
electron-electron interactions. First-principles calculations corroborate the 
experimental findings, predicting a magnetocrystalline anisotropy of 
0.987 MJ/m$^3$ (0.287 meV/f.u.) and dynamical stability via phonon analysis. 
To elucidate the finite-size effects, additional DFT calculations were performed 
on nanoclusters, which reveal significant surface-driven deviations 
from bulk properties, including enhanced magnetic moment and spin polarization 
in larger clusters. The combination of high magnetic performance, thermal 
robustness, and tunable nanoscale effects in Ni$_2$FeAl make it
as a promising candidate for wide range of applications, including 
nano-spintronics, high-density magnetic recording, magnetocaloric cooling, 
and magnetic sensing technologies.

\section{Acknowledgments}

PY thanks the Ministry of Education (MoE) for the fellowship and expresses sincere gratitude towards Dr. Ajay Kumar and Dr. Jyotsana Kala for useful discussion. We acknowledge the Department of Physics, IIT Delhi for providing XRD, PPMS and MPMS facilities and central research facility (CRF), IIT Delhi for FE-SEM, and HR-TEM measurements. PARAM Rudra, a national supercomputing facility at Inter-University Accelerator Centre (IUAC), New Delhi, is gratefully acknowledged for the theoretical results. RSD acknowledges the BRNS for financial support through DAE Young Scientist Research Award with Project Sanction No. 34/20/12/2015/BRNS. BKM acknowledges the funding support from SERB, DST (CRG/2022/000178).




\begin{thebibliography}{99}

\bibitem{Krenke_PRB_2006} T. Krenke, M. Acet, E. F. Wassermann, X. Moya, L. Manosa, and A. Planes, Ferromagnetism in the austenitic and martensitic states of Ni–Mn–In alloys, Phys. Rev. B {\bf73}, 174413 (2006).

\bibitem{Wang_JAP_2007} Y. D. Wang, Y. Ren, Z. H. Nie, D. M. Liu, L. Zuo, H. Choo, H. Li, P. K. Liaw, J. Q. Yan, R. J. McQueeney, J. W. Richardson, and A. Huq, Structural transition of ferromagnetic Ni$_2$MnGa nanoparticles, J. of Appl. Phys. {\bf101}, 063530 (2007).

\bibitem{Krenke_Nat_2005} T. Krenke, E. Duman, M. Acet, E. F. Wassermann, X. Moya, L. Manosa and A. Planes, Inverse magnetocaloric effect in ferromagnetic Ni–Mn–Sn alloys, Nat. Mater. {\bf4}, 450 (2005).

\bibitem{Groot_PRL_1983} R. A. de Groot, F. M. Mueller, P. G. van Engen, and K. H. J. Buschow, New class of materials: half-metallic ferromagnets, Phys. Rev. Lett. {\bf50}, 2024 (1983).

\bibitem{PY_JMMM_2025} P. Yadav, B.K. Mani, R.S. Dhaka, Exploring the role of disorder in Griffith-like magnetic phase transition in Co$_2$TiSi$_{0.5}$Al$_{0.5}$ Heusler alloy for magnetocaloric and spintronic applications, J. Magn. Magn. Mat. {\bf628}, 173093 (2025).

\bibitem{Galanakis_JPCM_2002} I. Galanakis, Surface properties of the half-and full-Heusler alloys, J. Phys.: Condens. Matter {\bf14}, 6329–6340  (2002).

\bibitem{Endo_JPDP_2011} H. Endo, A. Hirohata, J. Sagar, L.R. Fleet, T. Nakayama, K. O’Grady, Effect of grain size on exchange-biased Heusler alloys, J. Phys. D: Appl. Phys. {\bf44}, 345003 (2011).

\bibitem{Hashemifar_PRL_2005} S.J. Hashemifar, P. Kratzer, M. Scheffler, Preserving the Half-Metallicity at the Heusler Alloy Co$_2$MnSi (001) Surface: A Density Functional Theory Study, Phys. Rev. Lett. {\bf94}, 096402 (2005).

\bibitem{Wang_PRB_2005} W.H. Wang, M. Przybylski, W. Kuch, L.I. Chelaru, J. Wang, Y.F. Lu, J. Barthel, H.L. Meyerheim, J. Kirschner, Magnetic properties and spin polarization of Co$_2$MnSi Heusler alloy thin films epitaxially grown on GaAs(001), Phys. Rev. B {\bf71}, 144416 (2005).

\bibitem{Bosu_PRB_2011} S. Bosu, Y. Sakuraba, K. Uchida, K. Saito, T. Ota, E. Saitoh, K. Takanashi, Spin Seebeck effect in thin films of the Heusler compound Co$_2$MnSi, Phys. Rev. B {\bf83}, 224401 (2011).

\bibitem{Galdun_ACSN_2018} L. Galdun, V. Vega, Z. Vargová, E.D. Barriga-Castro, C. Luna, R. Varga, V.M. Prida, Intermetallic Co$_2$FeIn Heusler Alloy Nanowires for Spintronics Applications, ACS Appl. Nano Mater. {\bf1}, 7066–7074 (2018).

\bibitem{Karim_Nanot_2022} M.R. Karim, S. Naryan Panda, A. Barman, I. Sarkar, Strain and crystallite size controlled ordering of Heusler nanoparticles having high heating rate for magneto-thermal application, Nanotechnology {\bf33}, 235701  (2022).

\bibitem{GD_ACS_2021} U.R. Dahiya, G.D. Gupt, R.S. Dhaka, D. Kalyanasundaram, Functionalized Co$_2$FeAl Nanoparticles for Detection of SARS CoV-2 Based on Reverse Transcriptase Loop-Mediated Isothermal Amplification, ACS Appl. Nano Mater. {\bf4}, 5871–5882 (2021).

\bibitem{PRL_2007}G. Rollmann, M.E. Gruner, A. Hucht, R. Meyer, P. Entel, M.L. Tiago, J.R. Chelikowsky, Shellwise Mackay Transformation in Iron Nanoclusters, Phys. Rev. Lett. {\bf99}, 083402 (2007).  

\bibitem{Sun_sci_2000}S. Sun, C.B. Murray, D. Weller, L. Folks, A. Moser, Monodisperse FePt Nanoparticles and Ferromagnetic FePt Nanocrystal Superlattices, Science {\bf287}, 1989–1992 (2000).

\bibitem{Poorvi_EJIC_2024} P. Gupta, P. Yadav, B. Rana, B. Ghosh, R.S. Dhaka, K. Manna, Ni$_2$MnAl Heusler Alloy Nanoparticles for Chemoselective Hydrogenation of Nitro Compounds, Eur. J. Inorg. Chem., {\bf27} e202400157 (2024).

\bibitem{Billas_JMMM_97}I.M.L. Billas, A. Châtelain, W.A. De Heer, Magnetism of Fe, Co and Ni clusters in molecular beams, J. Magn. Magn. Mat. {\bf168}, 64–84 (1997).

\bibitem{Fabris_Nano_2019} F. Fabris, E. Lima, E. De Biasi, H.E. Troiani, M. Vásquez Mansilla, T.E. Torres, R. Fernández Pacheco, M.R. Ibarra, G.F. Goya, R.D. Zysler, E.L. Winkler, Controlling the dominant magnetic relaxation mechanisms for magnetic hyperthermia in bimagnetic core–shell nanoparticles, Nanoscale {\bf11}, 3164–3172 (2019).

\bibitem{Nehla_JALCOM_2019} P. Nehla, C. Ulrich, R. S. Dhaka, Investigation of the structural, electronic, transport and magnetic properties of Co$_2$FeGa Heusler alloy nanoparticles, J. Alloys Compds. {\bf776}, 379–386 (2019).

\bibitem{Ahmad_JMMM_2019} A. Ahmad, S. Mitra, S. K. Srivastava and A. K. Das, Size-dependent structural and magnetic properties of disordered Co$_2$FeAl Heusler alloy nanoparticles, J. Magn. Magn. Mater. {\bf474}, 599 (2019).

\bibitem{Xu_AIP_2018} Y. Xu, D. Yang, Z. Luo, F. Wu, C. Chen, M. Liu, L. Yi, H.-G. Piao, G. Yu, Fabrication and magnetic properties of structure-tunable Co$_2$FeGa-SiO$_2$ Heusler nanocompounds, AIP Advances {\bf8}, 055107 (2018). 

\bibitem{Wang_JMCC_2016} C. Wang, A.A. Levin, S. Fabbrici, L. Nasi, J. Karel, J. Qian, C.E. Viol Barbosa, S. Ouardi, F. Albertini, W. Schnelle, H. Borrmann, G.H. Fecher, C. Felser, Tunable structural and magnetic properties of chemically synthesized dual-phase Co$_2$NiGa nanoparticles, J. Mater. Chem. C., {\bf4} 7241–7252, (2016).

\bibitem{Wollmann_PRB_2015} L. Wollmann, S. Chadov, J. Kübler, and C. Felser, Magnetism in tetragonal manganese-rich Heusler compounds, Phys. Rev. B {\bf92}, 064417 (2015).

\bibitem{Webster_PM_1984} P. J. Webster, K. R. A. Ziebeck, S. L. Town and M. S. Peak, Magnetic order and phase transformation in Ni$_2$MnGa, Philosophical Magazine B, {\bf49}, 295–310 (1984).

\bibitem{Uijttewaal_PRL_2009} M. A. Uijttewaal, T. Hickel, J. Neugebauer, M. E. Gruner, and P. Entel, Understanding the Phase Transitions of the Ni$_2$MnGa Magnetic Shape Memory System from First Principles, Phys. Rev. Lett. {\bf102}, 035702 (2009).

\bibitem{Winterlik_AM_2013} J. Winterlik, S. Chadov, A. Gupta, V. Alijani, T. Gasi, K. Filsinger, B. Balke, G.H. Fecher, C. A. Jenkins, F. Casper, J. Kübler, G. Liu, L. Gao, S. S. P. Parkin and C. Felser, Design Scheme of New Tetragonal Heusler Compounds for Spin‐Transfer Torque Applications and its Experimental Realization, Adv. Mater. {\bf24}, 6283–6287 (2012).

\bibitem{Wu_APL_2009} F. Wu, S. Mizukami, D. Watanabe, H. Naganuma, M. Oogane, Y. Ando, and T. Miyazaki, Epitaxial Mn$_{2.5}$Ga thin films with giant perpendicular magnetic anisotropy for spintronic devices, Applied Physics Letters 94 (2009) 122503.

\bibitem{Zhang_JPCM_2007} W. Zhang, Z. Qian, J. Tang, L. Zhao, Y. Sui, H. Wang, Y. Li, W. Su, M. Zhang, Z. Liu, G. Liu, and G. Wu, Superparamagnetic behaviour in melt-spun Ni$_2$FeAl ribbons, J. Phys.: Condens. Matter {\bf19}, 096214 (2007).

\bibitem{Slebarski_PRB_2001} A. Ślebarski, M. B. Maple, A. Wrona, and A. Winiarska, Kondo-type behavior in Fe$_{2-x}$M$_x$TiSn (M=Co,Ni), Phys. Rev. B {\bf63}, 214416 (2001).

\bibitem{Slebarski_PRB_2002} A. Ślebarski, A. Wrona, T. Zawada, A. Jezierski, A. Zygmunt, K. Szot, S. Chiuzbaian, and M. Neumann, Electronic structure of some Heusler alloys based on aluminum and tin, Phys. Rev. B {\bf65}, 144430 (2002).

\bibitem{SI} See Supplementary Information for Experimental section: sample synthesis, FESEM-EDS results, additional magnetic and transport analysis. Theoretical section: Orbital resolved DOS for bulk-Ni$_2$FeAl and pair-polarization index calculations.

\bibitem{Kresse_PRB_96} G. Kresse and J. $\text{Furthm\"{u}ller}$, efficient iterative schemes for ab initio total-energy calculations using a plane-wave basis set, Phys. Rev. B {\bf54}, 11169 (1996).

\bibitem{Kresse_PRB_99} G. Kresse and D. Joubert, From Ultrasoft pseudopotentials to the projector augmented-wave method, Phys. Rev. B {\bf59}, 1758 (1999).

\bibitem{Monkhorst_PRB_76} H.J. Monkhorst and J. D. Pack, Special points for Brillouin-zone integrations, Phys. Rev. B {\bf13}, 5188 (1976).

\bibitem{Togo_SM_2015} A. Togo and I. Tanaka, First principles phonon clculations in materials science, Scripta Materialia {\bf108}, 1 (2015).

\bibitem{Read_JMMM_1984} D. A. Read, T. Moyo and G. C. Hallam, Low temperature magnetic hardness of melt spun Fe-Zr amorphous alloys, J. Magn. Magn. Mat. {\bf44}, 279–286 (1984).

\bibitem{Ajay_PRB_2024} A. Kumar, B. Schwarz, and R. S. Dhaka, Correlation between the exchange bias effect and antisite disorder in Sr$_{2-x}$La$_x$CoNbO$_6$(x= 0,0.2), Phys. Rev. B {\bf109}, 104434 (2024). 

\bibitem{Gabay_PRL_1981} M. Gabay, G. Toulouse, Coexistence of Spin-Glass and Ferromagnetic Orderings, Phys. Rev. Lett. {\bf47}, 201–204 (1981).

\bibitem{Anand_PRB_2012} V. K. Anand, D. T. Adroja, and A. D. Hillier, Ferromagnetic cluster spin-glass behavior in PrRhSn$_3$, Phys. Rev. B {\bf85}, 014418 (2012).

\bibitem{Jonsson_ACL_2003} P. E. Jonsson, Superparamagnetism and spin glass dynamics of interacting magnetic nanoparticle systems. Adv Chem Phys, vol. 128, Wiley Online Library; 2003, p.191–248.

\bibitem{Kittel_RMP_1949} C. Kittel, Physical Theory of Ferromagnetic Domains, Rev. Mod. Phys. {\bf21}, 541–583 (1949). 

\bibitem{Deka_JMMM_2016} B. Deka, R. Modak, P. Paul, and A. Srinivasan, Effect of atomic disorder on magnetization and half-metallic character of Cr$_2$CoGa alloy, J. Magn. Magn. Mat. {\bf418}, 107–111 (2016).
 
\bibitem{Srivastava_PCCP_2024} M. Srivastava, G. S. Bisht, and A. Srinivasan, Single-domain Fe$_2$CoGa$_{0.5}$Al$_{0.5}$ Heusler alloy nanaoparticles with enhanced properties, Phys. Chem. Chem. Phys. {\bf26}, 2863 (2024).

\bibitem{Cardona_ApplSci_2019} F. Arteaga-Cardona, N. G. Martha-Aguilar, J. O. Estevez, U. Pal, M. Á. Méndez-Rojas, and U. Salazar-Kuri, Variations in magnetic properties caused by size dispersion and particle aggregation on CoFe$_2$O$_4$, SN Appl. Sci. {\bf1}, 412 (2019).

\bibitem{Paswan_JPCS_2021} S. K. Paswan, S. Kumari, M. Kar, A. Singh, H. Pathak, J. P. Borah, and L. Kumar, Optimization of structure-property relationships in nickel ferrite nanoparticles annealed at different temperature, J. Phys. Chem. Solids {\bf151}, 109928 (2021). 

\bibitem{Trudel_JPDAP_2010} S. Trudel, O. Gaier, J. Hamrle, and B. Hillebrands, Magnetic anisotropy, exchange and damping in cobalt-based full-Heusler compounds: an experimental review, J. Phys. D: Appl. Phys. {\bf43}, 193001 (2010).

\bibitem{Kubota_JAP_2009} T. Kubota, J. Hamrle, Y. Sakuraba, O. Gaier, M. Oogane, A. Sakuma, B. Hillebrands, K. Takanashi, and Y. Ando, Structure, exchange stiffness, and magnetic anisotropy of Co$_2$MnAl$_x$Si$_{1−x}$ Heusler compounds, J. Appl. Phys. {\bf106} 113907 (2009).

\bibitem{Ritchie_PRB_2003} L. Ritchie, G. Xiao, Y. Ji, T. Y. Chen, C. L. Chien, M. Zhang, J. Chen, Z. Liu, G. Wu, and X. X. Zhang, Magnetic, structural, and transport properties of the Heusler alloys Co$_2$MnSi and NiMnSb, Phys. Rev. B {\bf68}, 104430 (2003).

\bibitem{Kittel} C. Kittel, Introduction to solid state physics, 7. ed, Wiley, New York, NY, 1996.

\bibitem{Gaier_JAP_2008} O. Gaier, J. Hamrle, S. J. Hermsdoerfer, H. Schultheiß, B. Hillebrands, Y. Sakuraba, M. Oogane, and Y. Ando, Influence of the L2$_1$ ordering degree on the magnetic properties of Co$_2$MnSi Heusler films, J. Appl. Phys. {\bf103}, 103910 (2008).

\bibitem{Kubler_PRB_2007} J. Kübler, G. H. Fecher, and C. Felser, Understanding the trend in the Curie temperatures of Co$_2$-based Heusler compounds: Ab initio calculations, Phys. Rev. B {\bf76}, 024414 (2007). 

\bibitem{Oogane_JAP_2009} M. Oogane, M. Shinano, Y. Sakuraba, and Y. Ando, Tunnel magnetoresistance effect in magnetic tunnel junctions using epitaxial Co$_2$FeSi Heusler alloy electrode, J. Appl. Phys. {\bf105}, 07C903 (2009).

\bibitem{Poddar_JMMM_2006} P. Poddar, J. Gass, D. J. Rebar, S. Srinath, H. Srikanth, S. A. Morrison, and E. E. Carpenter, Magnetocaloric effect in ferrite nanoparticles, J. Magn. Magn. Mat. {\bf307}, 227–231 (2006).

\bibitem{Michael_JMMM_1992} R. D. McMichael, R. D. Shull, L. J. Swartzendruber, L. H. Bennett, and R. E. Watson, Magnetocaloric effect in superparamagnets, J. Magn. Magn. Mat. {\bf111}, 29–33 (1992). 

\bibitem{PriyankaPRB} P. Nehla, Y. Kareri, G. D. Gupt, J. Hester, P. D. Babu, C. Ulrich, and R. S. Dhaka, Neutron diffraction and magnetic properties of Co$_2$Cr$_{1-x}$Ti$_x$Al Heusler alloys, Phys. Rev. B, {\bf 100}, 144444 (2019). 

\bibitem{PriyankaJAP} P. Nehla, V. K. Anand, B. Klemke, B. Lake, and R. S. Dhaka, Magnetocaloric properties and critical behavior of Co$_2$Cr$_{1-x}$Mn$_x$Al Heusler alloys, J. Appl. Phys., {\bf 126}, 203903 (2019). 

\bibitem{Ahmad_JMMM_2021} A. Ahmad, S. Mitra, S. K. Srivastava, and A. K. Das, Structural, magnetic, and magnetocaloric properties of Fe$_2$CoAl Heusler nanoalloy, J. Magn. Magn. Mat. {\bf540}, 168449 (2021). 

\bibitem{Banerjee_PL_1964} B. K. Banerjee, On a Generalised Approach to First and Second Order Magnetic Transitions, Phys. Lett. {\bf12}, 16 (1964).

\bibitem{Kaul_JMMM_85} S. N. Kaul, Static critical phenomena in ferromagnets with quenched disorder, J. Magn. Magn. Mater. {\bf53}, 5-53 (1985).

\bibitem{Tong_PRB_2020} S. Tong, X. Zhao, D. Wei, and J. Zhao, Low-temperature resistivity anomaly and weak spin disorder in Co$_2$MnGa epitaxial thin films, Phys. Rev. B {\bf101}, 184434 (2020).

\bibitem{Saha_PRB_2023} P. Saha, M. Singh, V. Nagpal, P. Das, and S. Patnaik, Scaling analysis of anomalous Hall resistivity and magnetoresistance in the quasi-two-dimensional ferromagnet Fe$_3$GeTe$_2$, Phys. Rev. B {\bf107}, 035115 (2023).

\bibitem{Madduri_PRB_2017} P. V. P. Madduri, and S. N. Kaul, Magnon-induced interband spin-flip scattering contribution to resistivity and magnetoresistance in a nanocrystalline itinerant-electron ferromagnet: Effect of crystallite size, Phys. Rev. B {\bf95}, 184402 (2017).

\bibitem{Helman_PRL_1976} J. S. Helman, and B. Abeles, Tunneling of Spin-Polarized Electrons and Magnetoresistance in Granular Ni Films, Phys. Rev. Lett. {\bf37}, 1429–1432 (1976).

\bibitem{Kondo_PTP_1964} J. Kondo, Resistance Minimum in Dilute Magnetic Alloys, Prog. Theor. Phys. {\bf32}, 37–49 (1964).

\bibitem{Bergmann_PR_1984} G. Bergmann, Weak localization in thin films, Physics Reports {\bf107}, 1–58 (1984).

\bibitem{Lee_RMP_1985} P. A. Lee, and T. V. Ramakrishnan, Disordered electronic systems, Rev. Mod. Phys. {\bf57}, 287–337 (1985).

\bibitem{Dhara_PRB_2016} S. Dhara, R. R. Chowdhury, and B. Bandyopadhyay, Observation of resistivity minimum at low temperature in Co$_x$Cu$_{1-x}$(x$\sim$0.17-0.76) nanostructured granular alloys, Phys. Rev. B {\bf93}, 214413 (2016).

\bibitem{Fabietti_PRB_2010} L. M. Fabietti, J. Ferreyra, M. Villafuerte, S. E. Urreta, and S. P. Heluani, Kondo-like effect in magnetoresistive CuCo alloys, Phys. Rev. B {\bf82}, 172410 (2010).

\bibitem{Mei_PRB_1986} Y. Mei, and H. L. Luo, Resistivity minima in Au$_{1-x}$Ni$_x$ alloys (0.30$\leqslant$x$\leqslant$0.42), Phys. Rev. B {\bf34}, 509–514 (1986).

\bibitem{Muthuselvam_JALCOM_2012} I. P. Muthuselvam, and R. N. Bhowmik, Grain size dependent magnetization, electrical resistivity and magnetoresistance in mechanically milled La$_{0.67}$Sr$_{0.33}$MnO$_3$, J. Alloys Compds. {\bf511}, 22–30 (2012). 

\bibitem{Young_JAP_2014} J. C. Prestigiacomo, D. P. Young, P. W. Adams, and S. Stadler, Hall effect and the magnetotransport properties of Co$_2$MnSi$_{1-x}$Al$_x$ Heusler alloys, J. Appl. Phys. {\bf115}, 043712 (2014). 

\bibitem{Wen_I_2018} Z. Wen, H. Hou, J. Tian, Y. Zhao, H. Li, and P. Han, First-principles investigation of martensitic transformation and magnetic properties of Ni$_2$XAl (X=Cr, Fe, Co) Heusler compounds, Intermetallics {\bf92}, 15–19 (2018).

\bibitem{Sehoul_JSNM_2025} B. Sehoul, T. Das, A. Dutta, T. I. Al-Muhimeed, and S. H. A. Ahmad, Compressive Evaluation of Structural, Electronic, Elastic, and Magnetic Features of Ni$_2$XAl (X = V, Fe) Heusler Alloys: A DFT Insight, J. Supercond. Nov. Magn. {\bf38}, 1 (2025). 

\bibitem{Benhizia_CCM_20} N. E. Benhizia, Y. Zaoui, S. Amari, L. Beldi, and B. Bouhafs, Theoretical study of structural, electronic, dynamic and thermodynamic properties of Ni$_2$FeAl and Ni$_2$CoAl alloys, Computational Condensed Matter {\bf24}, e00480 (2020).

\bibitem{Qawasmeh_JAP_2012} Y. Qawasmeh, and B. Hamad, Investigation of the structural, electronic, and magnetic properties of Ni-based Heusler alloys from first principles, J. Appl. Phys. {\bf111}, 033905 (2012).

\bibitem{Herper_PRB_2018} H. C. Herper, Ni-based Heusler compounds: How to tune the magnetocrystalline anisotropy, Phys. Rev. B {\bf98}, 014411 (2018).


\bibitem {Parkin_PRM_2017} S. V. Faleev, Y. Ferrante, J. Jeong, M. G. Samant, B. Jones, and S. S. P. Parkin, Heusler compounds with perpendicular magnetic anisotropy and large tunneling magnetoresistance, Phys. Rev. Materials {\bf1}, 024402 (2017). 

\bibitem{Bruno_PRB_1989} P. Bruno, Tight-binding approach to the orbital magnetic moment and magnetocrystalline anisotropy of transition-metal monolayers, Phys. Rev. B {\bf39}, 865–868 (1989). 

\bibitem{Umetsu_APL_2006} R. Y. Umetsu, A. Sakuma, and K. Fukamichi, Magnetic anisotropy energy of antiferromagnetic L1-type equiatomic Mn alloys, Applied Physics Letters {\bf89}, 052504 (2006).

\bibitem{vleck_PRB_1937} J. H. Van Vleck, On the Anisotropy of Cubic Ferromagnetic Crystals, Phys. Rev. {\bf52}, 1178–1198 (1937).

\bibitem{Kubler_PRB_1983} J. Kübler, A. R. William, and C. B. Sommers, Formation and coupling of magnetic moments in Heusler alloys, Phys. Rev. B {\bf28}, 1745–1755 (1983).

\bibitem{Zayak_PRB_2008} A. T. Zayak, P. Entel, and J. R. Chelikowsky, Minority-spin polarization and surface magnetic enhancement in Heusler clusters, Phys. Rev. B {\bf77}, 212401 (2008).

\bibitem{Zayak_JAP_2008} A. T. Zayak, S. P. Beckman, M. L. Tiago, P. Entel, and J. R. Chelikowsky, Switchable Ni–Mn–Ga Heusler nanocrystals, J. Appl. Phys. {\bf104}, 074307 (2008). 

\end{thebibliography}
\end{document}